\documentclass[epj,nopacs]{svjour}

\usepackage{dcolumn}
\usepackage{graphicx}
\usepackage{amsmath}
\usepackage{amssymb}
\usepackage{bm}
\usepackage{revsymb}

\usepackage{bm}

\date{August 7, 2009}

\begin{document}


\title{Nuclear form factor, validity of the equivalent
photon approximation and Coulomb corrections to muon pair\\
production in photon-nucleus and nucleus-nucleus
collisions}

\newcommand{\addrROLLA}{Department of Physics,
Missouri University of Science and Technology,
Rolla, Missouri 65409-0640, USA}

\newcommand{\addrHDphiltheo}{Institut f\"ur Theoretische Physik,
Universit\"{a}t Heidelberg,
Philosophenweg 16, 69120 Heidelberg, Germany}

\newcommand{\addrNOVOSIBIRSK}{Novosibirsk State University,
Pirogova 2, 630090, Novosibirsk, Russia}

\authorrunning{U. D. Jentschura, V. G. Serbo}
\titlerunning{Muon Pair Production}

\author{U.D.~Jentschura \inst{1,2},$\,$
V.G.~Serbo \inst{2,3,}\thanks{Corresponding author:
serbo@math.nsc.ru}}

\institute{{\addrROLLA} \and {\addrHDphiltheo} \and
{\addrNOVOSIBIRSK}}

\abstract{We study in detail the influence of the nuclear form factor
both on the Born cross section and on the Coulomb
corrections to the photo-production of muon pairs off heavy
nuclei ($\gamma Z \to  \mu^+ \mu^- Z$) and in heavy-ion
collisions ($ZZ \to ZZ \mu^+ \mu^-$). Our findings indicate
a number of issues which have not been sufficiently
described as yet in the literature: {\em (i)} the use of a
realistic form factor, based on the Fermi charge
distribution for the nucleus, is absolutely indispensable
for reliable theoretical predictions; {\em (ii)} we checked
quantitatively that the equivalent photon approximation has
a very good accuracy for the discussed processes; and {\em
(iii)}  we present a leading logarithmic calculation of the
Coulomb corrections which correspond to multi-photon
exchange of the produced $\mu^{\pm}$ with the nuclei. These
corrections are found to be small (on the percent level).
Our result justifies
using the Born approximation for numerical simulations of
the discussed process at the RHIC and LHC colliders.
Finally, we calculate the total cross section for muon pair
production at RHIC and LHC.}


\maketitle

%
%
%
\section{Introduction}

Lepton pair production in ultra-relativistic nuclear
collisions was discussed in numerous papers
(see~\cite{BHTSK-2002,BHT-2007,PhysRev-2008} for a review
and references therein). For definiteness, we restrict
ourselves to equal charge numbers of the nuclei
$Z_1=Z_2\equiv Z$ and symmetric Lorentz factors
$\gamma_1=\gamma_2\equiv \gamma$, for the RHIC and the LHC
colliders with parameters given in Table~\ref{t1}.

\begin{figure}[tb]
\begin{center}
\includegraphics[width=0.6\linewidth]{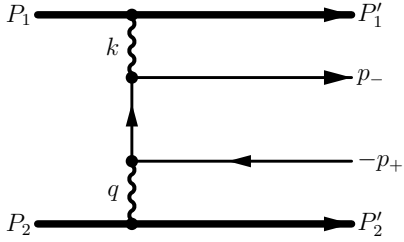}
\caption{\label{fig1} Feynman diagram for the lepton pair
production $ZZ\to ZZl^+l^-$in the Born approximation
($l=e,\;\mu$) }
\end{center}
\end{figure}

In the present paper, we primarily consider the production
of a muon pair, but for completeness and comparison, we
first recall some results for electron-positron ($e^+e^-$)
pair production and therefore make a slight detour. The
production of a single $e^+e^-$ pair in the Born
approximation is described by the Feynman diagram of
Fig.~\ref{fig1}; the corresponding cross section was
obtained many years ago~\cite{Landau}. Since the Born cross
section $\sigma^{e^+e^-}_{\rm Born}$ is huge (see
Table~\ref{t1}), the $e^+e^-$  pair production can be a
serious background for many experiments. It is also an
important issue for the beam lifetime and luminosity of
these colliders \cite{Klein:2000ba}. This means that
various corrections to the Born cross section, as well as
the cross section for $n$-pair production, are of great
importance. The subject is inherently difficult; a number
of controversial and incorrect statements in the literature
have been clarified in
Refs.~\cite{BHTSK-2002,ISS-1999,LM-2000,LMS-2002,JHS-2008,LM-2009}.

\begin{table}[!h]
\vspace{5mm} \caption{Cross sections for the production of
light lepton pairs at modern colliders}
\begin{center}
\begin{tabular}{c@{\hspace{0.5cm}}c@{\hspace{0.3cm}}c@{\hspace{0.3cm}}c@{\hspace{0.3cm}}c}
\hline \rule[-3mm]{0mm}{5mm}
Collider & $Z$ & $\gamma$ &
$\sigma^{e^+e^-}_{\rm
Born}$ [kb] & $\sigma^{\mu^+\mu^-}_{\rm Born}$ [b] \\
\hline \rule[-3mm]{0mm}{8mm}
RHIC, Au-Au & 79 & 108 & 36.0 & 0.209
\\ \rule[-3mm]{0mm}{5mm}
LHC, Pb-Pb & 82 & 3000 & 227 & 2.46 \\
\hline
\end{tabular}
\label{t1}
\end{center}
\end{table}

Since the parameter  $Z\alpha$ is not small ($Z\alpha
\approx 0.6$ for Au-Au and Pb-Pb collisions), the whole
series in $Z\alpha$ has to be summed in order to obtain the
cross section with sufficient accuracy unless higher-order
corrections are otherwise
parametrically suppressed. The exact cross section for single
pair production $\sigma_1$ can be represented as the sum of
the Born value, the Coulomb correction, and of the
unitarity correction,
\begin{equation}
\sigma_1 = \sigma_{\rm Born} + \sigma_{\rm Coul}+ \,
\sigma_{\rm unit}\,.
\label{1}
\end{equation}

%
%

%
%
\begin{figure}[tb]
\begin{center}
\includegraphics[width=0.6\linewidth]{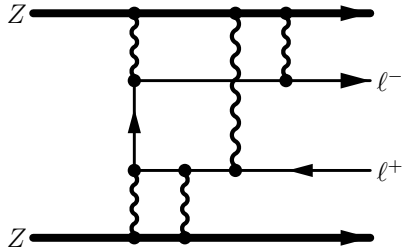}
\caption{\label{fig2} Feynman digram for the Coulomb
correction}
\end{center}
\end{figure}

\begin{figure}[thb]
\begin{center}
\includegraphics[width=0.9\linewidth]{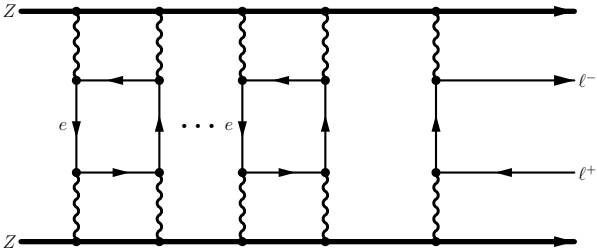}
\caption{Feynman diagram for the unitarity correction}
\label{fig3}
\end{center}
\end{figure}

The Coulomb correction $\sigma_{\rm Coul}$ corresponds to
multi-photon exchange of the produced $e^{\pm}$ with the
nuclei (Fig.~\ref{fig2}); it was calculated in
Ref.~\cite{ISS-1999,LM-2009}. The unitarity correction
$\sigma_{\rm unit}$ corresponds to the exchange of
light-by-light blocks between the nuclei (Fig.~\ref{fig3});
it was calculated in~\cite{LMS-2002,JHS-2008}. It was found
that the Coulomb corrections are about 10~\% while the
unitarity corrections are about two times smaller (see
Table~\ref{t2}). In the last column of Table~\ref{t2} is
shown the result of Baltz (see Ref.~\cite{Baltz-2005})
obtained by numerical calculations using formula for the
cross section resulting from ``exact solution of the
semiclassical Dirac equations.'' In fact, the employed formulas
allow to calculate the Coulomb correction in the leading
logarithmic approximation only, and this may account for the
discrepancies of the results for RHIC indicated in
Table~\ref{t2}. For the case of electron-positron pairs,
the leading logarithmic approximation is insufficient
because of the large absolute magnitude of the correction.

%
%

%
%
\begin{table*}[thb]
\caption{\label{t2} Coulomb and unitarity corrections to
the $e^+e^-$ pair production}
\begin{center}
\par
\begin{tabular}{c@{\hspace{0.5cm}}c@{\hspace{0.5cm}}c@{\hspace{0.5cm}}c@{\hspace{0.5cm}}c}
\hline
\rule[-3mm]{0mm}{8mm} Collider &
$\frac{\displaystyle\sigma_{\rm Coul}}{\displaystyle
\sigma_{\rm Born}}$ (Refs.~\cite{ISS-1999,LM-2009}) &
$\frac{\displaystyle\sigma_{\rm
unit}}{\displaystyle\sigma_{\rm Born}}$
(Refs.~\cite{LMS-2002,JHS-2008}) & $\frac{\displaystyle
\sigma_{\rm Coul}}{\displaystyle
\sigma_{\rm Born}}$ (Ref.~\cite{Baltz-2005}) \\
\hline
\rule[-3mm]{0mm}{8mm}
RHIC, Au-Au &  $-10$\% & $-5.0$\% &  $-17$\% \\
\rule[-3mm]{0mm}{5mm}
LHC, Pb-Pb & $-9.4$\% & $-4.0$\% & $-11$\% \\
\hline
\end{tabular}
\end{center}
\end{table*}

In this paper, we present detailed calculations related to
muon pair production. This process may be easier to
observe experimentally than $e^+e^-$ pair production.
It should be stressed that the calculational
scheme, as well as, the final results for the $\mu^+\mu^-$
pair production are quite different from those for the
$e^+e^-$ pair production.

The principal issues related to muon pair production,
including the problem of unitarity corrections, have been
considered in Refs.~\cite{JHS-2008,HKS-2007}. In
particular, using simple estimates, it was pointed out
that: {\it (i)} the Born contribution can be easily
calculated using the equivalent photon approximation (EPA)
which has in our particular case a good accuracy; {\it
(ii)} contrary to the $e^+e^-$ case, the Coulomb correction
is small in the muon case (on the level of a percent).
The last statement is of
principal importance because it justifies the
validity of the Born
approximation for event generators of this process at the
RHIC and LHC colliders.

In a recent paper~\cite{Baltz-2009}, the conclusion {\it (i)}
has been confirmed, but the point {\it (ii)} has been
questioned. Namely, in Ref.~\cite{Baltz-2009}, it was found out
that the Coulomb corrections to muon pair production are
rather large: $-22\,\%$ for RHIC and $-17\,\%$ for LHC.
These results have been obtained using the same formulas as
for the $e^+e^-$ case with the minor changes.
Below, we present a new calculation of the Coulomb
corrections for muon pair production in the leading
logarithmic approximation (LLA); our result is in
agreement with a previous numerically small estimate of the
Coulomb corrections as given in Ref.~\cite{HKS-2007},
but it is in strong disagreement with the result of
Ref.~\cite{Baltz-2009}.

We would like to note that the above features {\it (i)} and
{\it (ii)} are directly related to the fact that both the
electromagnetic form factors of the nuclei $F(K^2)$,
$F(Q^2)$ and the cross section for the virtual block
$\gamma^*(k)+\gamma^*(q) \to \mu^+\mu^-$ drop quickly with
increasing photon virtualities $K^2=-k^2>0$ and $Q^2=-q^2
\approx \bm{q}^2 >0$. However, the scale of this decrease
is much less for the nuclear form factor than for the
virtual $\gamma^*+\gamma^* \to \mu^+\mu^-$ block (by
$\gamma^*$, we here denote a virtual as opposed to a real
photon).

As a rule, the calculation of muon pair production for
nuclear collisions is very laborious (for example, the exact
expression for the Born cross section even for the case of
simplified form factors is an eight-fold integral).
Therefore, it is convenient to check the main points of various
approximations using the simpler process of muon
photo-production. In this case we have a possibility to
perform relatively easily both the exact and approximate
calculations and compare them.

This paper is organized as follows. In Sec.~\ref{section2},
we study in detail the photo-production of muon pair off
heavy nuclei $\gamma Z \to \mu^+ \mu^- Z$. An exact
calculation of the total Born cross section for arbitrary
photon energy starting from the threshold is carried out.
The use of a realistic form factor instead of simplified
form $F(Q^2)= 1/(1+Q^2/\Lambda^2)$ turns out to be
critically important for moderate photon energies. In
Sec.~\ref{section3}, the validity of the EPA is studied
both for a realistic and for a simplified representation of
the nuclear form factor. Coulomb corrections to
photo-production of the muon pair are studied in
Secs.~\ref{section4}. Predictions for the RHIC and LHC
colliders are given in Sec.~\ref{section5}, and we conclude
with a summary in Sec.~\ref{section6}. Throughout the
paper, we use a system of units in which $c=1$, $\hbar =1$,
$\alpha=e^2/(\hbar c) \approx 1/137$ and denote the muon
and nuclear mass as $m$ and $M$, respectively.

%
%
\section{Form factor and Born--level pair photo-production}
\label{section2}

\subsection{Form factors and nuclear charge distributions}

We first recall basic formulas related to the realistic and
simplified form factor representations for the colliding
heavy nuclei which are central to our investigation. For
the realistic form factor, we employ a Fermi-type
nuclear charge distribution in the form (see
Refs.~\cite{ADNDT})
\begin{equation}
\rho(r)=\frac{\rho_0}{1+\exp{[(r-R)/a]}}
\label{sff1}
\end{equation}
with $a=2.30/(4\ln{3})$ fm, $R=6.55$ fm for Au (mass number $A=197$) and
$R=6.647$ fm for Pb ($A=208$). This leads to the mean squared
radius
\begin{equation}
\sqrt{\left\langle r^2 \right\rangle}=\sqrt{\frac 35
\left[1+\frac 73 \left(\frac{\pi a}{R}\right)^2\right]}\,
R\,,
\end{equation}
with  $\sqrt{\left\langle r^2 \right\rangle}= 5.4338$ fm
for gold and $\sqrt{\left\langle r^2 \right\rangle}=
5.5041$ fm for lead. The latter numbers are in very good
agreement with the experimental values $\sqrt{\left\langle
r^2 \right\rangle}_{\rm exp}= 5.4358$ fm for gold and
$\sqrt{\left\langle r^2 \right\rangle}_{\rm exp}= 5.5010$
fm for lead.

\begin{figure}[tb]
\begin{center}
\includegraphics[width=1.0\linewidth]{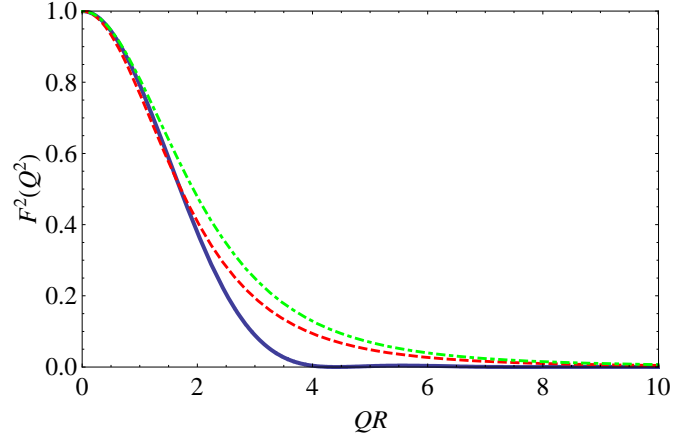}
\caption{Realistic (solid line) and simplified (dashed line
for $\Lambda=80$ MeV and dot-dashed line for $\Lambda=90$
MeV) form factors vs. $QR$ for Au}
 \label{F:4}
\end{center}
\end{figure}

The nuclear form factor is defined as
\begin{equation}
F({\bm q}^2) = \frac{1}{N}
\int \rho(r)\,{\rm e}^{-{\rm i}{\bm q} \cdot {\bm r}}\,
d^3r\,,\;\;N=\int\rho(r)\,d^3r \,,
\label{sff2}
\end{equation}
where ${\bm q}^2\approx Q^2$ and $\bm{q}$ is the
three-vector part of the photon four-momentum $q$. Its
behavior is shown on Fig.~\ref{F:4}, it is seen that for
$Q^2>1/R^2$, the form factor drops quickly with the growth
of $Q^2$. On the other hand, the cross section for the
virtual block $\gamma^*+\gamma^* \to \mu^+\mu^-$ drops
quickly with the growth of $Q^2$  at $Q^2> W^2=(k+q)^2>
(2m)^2$ [see Eq.~(\ref{gQ}) below]. For further
consideration it is important that
\begin{equation}
1/R^2\approx (30\;\mbox{ MeV})^2 \ll W^2\;.
\end{equation}

For {\it the simplified form factor}, we use an approximation of a
monopole form factor corresponding to an exponentially
decreasing charge distribution
\begin{equation}
\label{26}
F(Q^2) = \frac{1}{1+Q^2/\Lambda^2}\,.
\end{equation}
Its behavior is also shown in Fig.~\ref{F:4}. This
approximate form of the form factor is used, for example,
in Refs.~\cite{IM98,HKS-2007,Baltz-2009} and enables to
perform some calculations analytically. For the concrete
calculations reported in Refs.~\cite{HKS-2007,Baltz-2009},
the value
\begin{equation}
\Lambda=80\,{\rm MeV}
\end{equation}
is used for lead and gold. In the calculations below we
also use this value unless otherwise stated. Another
possibility is to use the connection of $\Lambda$ with the
mean squared radius $\sqrt{\left\langle r^2
\right\rangle}$, in this case
\begin{equation}
\Lambda = \sqrt{\frac{6}{\left\langle r^2
\right\rangle}}\approx 90 \;\mbox{MeV}
\end{equation}
for lead and gold.

Looking at three curves in Fig.~\ref{F:4}, one can come to
the conclusion that the difference between the two choices
of the $\Lambda$ parameter should be negligible. We will
show, however, that a transition from the realistic form
factor to the simplified one with $\Lambda = 80$ MeV or
$\Lambda = 90$ MeV results in a change of the total cross
section for the muon pair production at the RHIC collider
on the level of $10$ \% or $20$ \%, respectively.

%
%
\subsection{Realistic form factor: exact result for
the Born cross section}

\begin{figure}[htb]
\begin{center}
\includegraphics[width=0.7\linewidth]{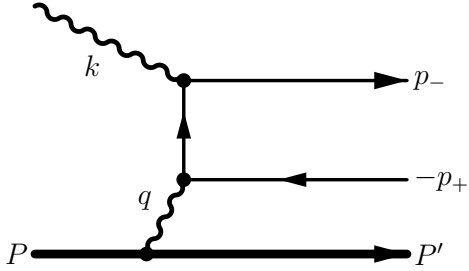}
\caption{\label{fig5} Feynman diagram for the
photo-production of muon pair in the Born approximation.
The incoming virtual photon has momentum $k$, the invariant
mass squared of the pair is $W^2 =(k +q)^2=(p_+ +p_-)^2$.
The four-momenta of the produced leptons are $p_\pm$}
\end{center}
\end{figure}

We start to discuss the role of the form factor on the
basis of muon pair production by a real photon with the
energy $\omega$ off the nucleus with charge $Ze$ and mass
$M$:
\begin{equation}
\gamma(k)+Z(P) \to \mu^+(p_+)+\mu^-(p_-)+ Z(P')\,.
\label{1process}
\end{equation}
In the Born approximation, this process is described by the
Feynman diagram of Fig.~\ref{fig5}, and the
corresponding diagram is also contained as a block diagram
within the Feynman diagram for pair production off
heavy nuclei. We assume that a real photon with 4-momentum
$k$ and a virtual photon with 4-momentum $q=P-P'$ and
virtuality
\begin{equation}
Q^2=-q^2 >0
\label{2}
\end{equation}
collide with each other and produce a {$\mu^{+}\mu^{-}$}
pair with the invariant mass squared
\begin{equation}
W^2 = (k +q)^2=2kq-Q^2\,.
\label{3}
\end{equation}
We also use the notations
\begin{equation}
\label{defomega}
 s=(k+P)^2 = M^2+2\omega M\,,\;\;\sigma_0 =\frac{Z^2
\alpha^3}{m^2} \,,
\end{equation}
where $m$ is the muon mass. So, $\omega$ measures the
incoming photon energy in the rest frame of the incoming
nucleus.

\begin{figure}[tb]
\begin{center}
\includegraphics[width=0.7\linewidth]{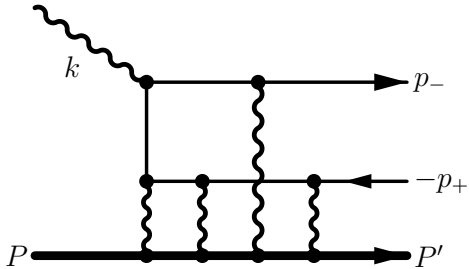}
\caption{Typical Feynman diagram for a higher-order Coulomb
correction to the photo-production of a muon pair}
 \label{fig6}
\end{center}
\end{figure}

The exact cross section for muon pair production
$\sigma_{\gamma Z}$ can be split into the form
\begin{equation}
\sigma_{\gamma Z} = \sigma_{\rm Born} + \sigma_{\rm
Coul}\,, \label{11master}
\end{equation}
where $\sigma_{\rm Born}$ corresponds to the Born cross
section, and the Coulomb correction $\sigma_{\rm Coul}$
corresponds to multi-photon exchange of the produced
$\mu^{\pm}$ with the nucleus (Fig. \ref{fig6}).

\begin{figure}[tb]
\begin{center}
\includegraphics[width=1.0\linewidth]{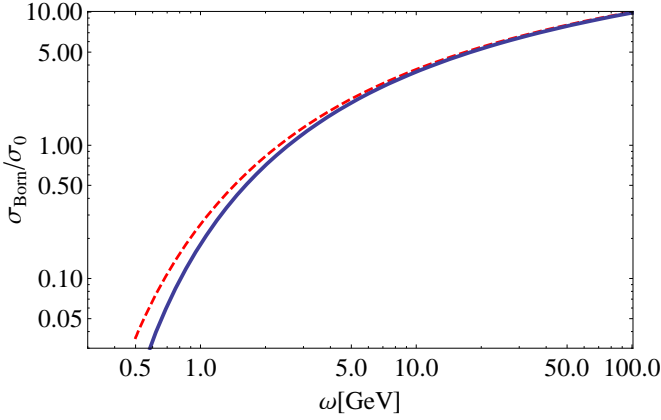}
\caption{\label{fig7} A Born cross section for the
realistic (solid line) and simplified (dashed line) form
factors (photo-production on Au)}
\end{center}
\end{figure}

It is well known (see, for example, Ref.~\cite{BGMS-75})
that the exact Born cross section for the process (\ref{1})
as well as for electro-production can be written in terms
of two structure functions or two cross sections
$\sigma_T(W^2, Q^2)$ and $\sigma_S(W^2, Q^2)$ for the
virtual processes $\gamma\gamma_T^* \to \mu^+\mu^-$ and
$\gamma\gamma_S^* \to \mu^+\mu^-$, respectively (here,
$\gamma$ is a real initial photon, while $\gamma_T^*$ and
$\gamma_S^*$ denote the virtual transverse and
scalar/longitudinal photons with helicity $\lambda_T=\pm 1$
and $\lambda_S=0$, respectively):
\begin{align}
d\sigma_{\rm Born} =& \; \sigma_T(W^2, Q^2)\, dn_T(W^2, Q^2)
\nonumber\\[2ex]
& \; + \sigma_S(W^2, Q^2)\,dn_S(W^2, Q^2) \,.
\label{39}
\end{align}
The coefficients $dn_T$ and $dn_S$ are called the number of
transverse and scalar virtual photons (generated by the
nucleus). The cross sections $\sigma_T$ and $\sigma_S$ can
be found in Appendix E of the review~\cite{BGMS-75}:
\begin{align}
\sigma_T  =& \; {4\pi \alpha ^2 \over W^2+Q^2} \Biggl\{ \left[ 1 +{4m^2
W^2 -8m^4 - 2Q^2 W^2\over (W^2+Q^2)^2} \right]\, L
\nonumber\\
& - \left[1
+{4m^2 W^2 - 4Q^2 W^2\over (W^2+Q^2)^2} \right]v\,\Biggr\}\,,
\label{sT} \\[2ex]
\sigma_S =& \; {16\pi \alpha^2 Q^2\,W^2\over (W^2+Q^2)^3} \left[v -
{2m^2\over W^2}\, L \right],
\label{sS}
\end{align}
where
\begin{equation}
\label{defvL}
v= \sqrt{1-{4m^2 \over W^2}}\,,\qquad
L= 2\ln{\left[{W\over 2m} (1+v)
\right]}\,.
\end{equation}
Let us note that
\begin{align}
\sigma_T \sim & \;  {4\pi \alpha ^2 \over W^2}\,
[1+{\cal O}(Q^2/W^2)]\,,\;
\nonumber\\[2ex]
\sigma_S\sim & \; {16\pi \alpha^2Q^2\over W^4}\;\;
\mbox{at} \;\;Q^2 \ll W^2\,,
\end{align}
and
\begin{equation}
\sigma_T  \sim {4\pi \alpha ^2 \over
Q^2}\,,\quad \sigma_S\sim {16\pi \alpha^2
W^2\over (Q^2)^2} \quad \mbox{for} \;\;Q^2 \gg W^2\,.
\label{gQ}
\end{equation}
The  number of photons can be found in Sec.~6 and
Appendix~D of Ref. \cite{BGMS-75}
\begin{eqnarray}
d n_T&=& \frac{Z^2\alpha}{\pi} \left( 1-y -\frac{M^2 y^2}{Q^2}\right) \,
F^2(Q^2)\, \frac{d W^2}{W^2+Q^2}\, \frac{d
Q^2}{Q^2}\,,
\nonumber \\
d n_S&=& \frac{Z^2\alpha}{\pi} \left( 1-y  +\frac{1}{4}y^2\right)F^2(Q^2)\,
\frac{d W^2}{W^2+Q^2}\,\frac{d Q^2}{Q^2}\,,
\label{numbers}
\end{eqnarray}
where
\begin{equation}
y=\frac{kq}{kP}=\frac{W^2+Q^2}{2\omega M}\,.
\label{number1}
\end{equation}
Integrating the cross section (\ref{39}) over $Q^2$ in the region
$Q^2_{\min} \leq Q^2 \leq Q^2_{\max}$, where (see Problem 3 to
\S~68 in Ref.~\cite{BLP})
\begin{align}
Q^2_{\min, \max} =& \; B\mp \sqrt{B^2-C}\,,
\\[2ex]
2B=& \; \frac{2M^6-M^4 W^2-M^2 W^4}{(2M^2+W^2)s}+s-2M^2-W^2
\nonumber\\[2ex]
\approx & \; 2\,\frac{2\omega^2-W^2(1-\omega/M)}{1+2\omega/M}\,,
\\[2ex]
C=& \; \frac{M^2 W^4}{s}\,,
\end{align}
and over $W$ in the region $2m\leq W \leq \omega$, we
obtain the exact result for the Born cross section
presented in Fig.~\ref{fig7}.

%
%
\subsection{Simplified form factor: exact result for the Born cross section}

The exact result for the Born cross section for the case of
a simplified form factor can be obtained using
Eqs.~(\ref{39}), (\ref{sT}), (\ref{sS}), (\ref{numbers})
with the form factor (\ref{26}). The result is shown by the
dashed line in Fig.~\ref{fig7}. It is seen that
calculations with the simplified form factor give an
accuracy better than 10~\%, 5~\% and 2~\% at $\omega > 3.5$
GeV, 8 GeV and 50 GeV, respectively. At RHIC, the region
near the ``accuracy threshold'' ($2m< \omega < 8$ GeV)
gives a numerically important contribution, which accounts
for about $10\div 20$ \% of the difference between cross
sections with the realistic and simplified form factors.

%
%
\section{Approximations to Born--level pair photo-production}
\label{section3}

%
%
\subsection{Realistic form factor: equivalent photon
approximation (EPA)}

Let us recall the usual schema of the EPA, but with the addition
of an accurate treatment of the nuclear form factor (see, for
example, Ref.~\cite{BGMS-75}). For the case of high-energy
photons $\omega\gg 2m$, the most important contribution to
the photo-production cross section stems from photons with
very small virtuality $Q^2 \ll {W}^2$ [we recall the
definition of $\omega$ in Eq.~(\ref{defomega}) and
that $Q^2 = -q^2 \approx \bm{q}^2$].  It means that we can
ignore the contribution of the scalar photons in
Eq.~(\ref{39}) and the dependence of $\sigma_T$ on $Q^2$;
besides we can simplify the expression for $dn_T$ from Eq.
(\ref{numbers}). As a result, we obtain the simple
approximate (EPA) expression
\begin{equation}
d\sigma^{\rm EPA}_{\rm Born} =
\sigma_{\gamma \gamma}(W^2)\, dn_\gamma(W^2, Q^2)\,,
\label{sEPA}
\end{equation}
where
\begin{align}
\sigma_{\gamma \gamma}(W^2) =& \;
\frac{4\pi \alpha ^2}{W^2}\left[ \left(1
+\frac{4m^2}{W^2} -\frac{8m^4}{W^4} \right)\, L
\right.
\nonumber\\[2ex]
& \; \left. - \left(1
+\frac{4m^2}{W^2}\right) v\,\right]\,,
\label{gg} \\[2ex]
d n_\gamma =& \;  \frac{Z^2\alpha}{\pi}
\left( 1- \frac{Q^2_{\min}}{Q^2}\right) \,
F^2(Q^2)\,
\frac{d W^2}{W^2}\,
\frac{d Q^2}{Q^2}\,,
\nonumber\\[2ex]
\qquad Q^2_{\min} =& \; \frac{W^4}{4\omega^2}\,.
\label{ng}
\end{align}
The quantities $v$ and $L$ are defined in Eq.~(\ref{defvL}).

\begin{figure}[tb]
\begin{center}
\includegraphics[width=1.0\linewidth]{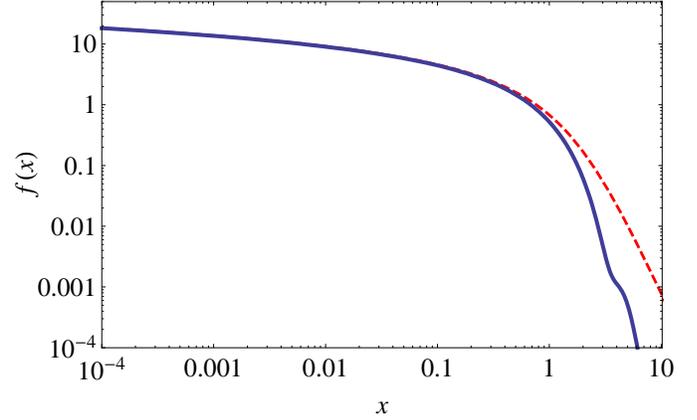}
\caption{The function $f(x)$ from Eq.~(\ref{28})  for the
realistic form factor (solid line) and
$\tilde{f}(x/(R\Lambda))$ from Eq.~(\ref{fsimff}) for the
simplified form factor (dashed line) (photo-production on
Au)}
 \label{fig8}
\end{center}
\end{figure}

\begin{figure}[t!]
\begin{center}
\includegraphics[width=1.0\linewidth]{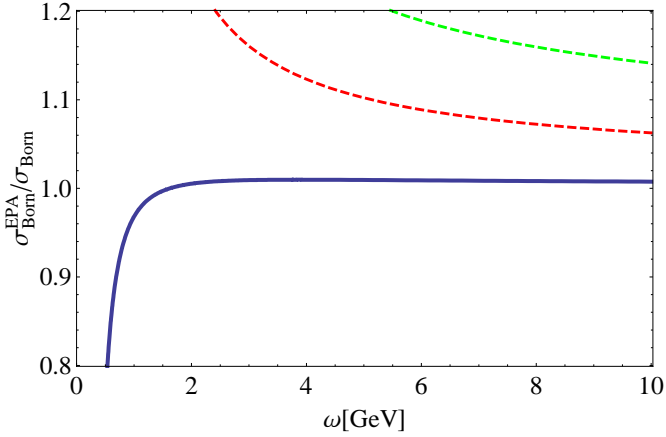}
\caption{Ratio $\sigma^{\rm EPA}_{\rm Born}/\sigma_{\rm
Born}$ (solid line) and $\sigma^{\rm SEPA}_{\rm
Born}/\sigma_{\rm Born}$ (dashed and dash-dotted lines),
where $\sigma_{\rm Born}$ is the total Born cross section
from Eq.~(\ref{39}) while $\sigma^{\rm EPA}_{\rm Born}$
from Eq.~(\ref{A6}) is calculated for the case of realistic
form factor and $\sigma^{\rm SEPA}_{\rm Born}$ from
Eq.~(\ref{A6s}) is calculated for the case of a simplified
form factor with $\Lambda =80$ MeV (dashed line) and
$\Lambda =90$ MeV (dot-dashed line) (photo-production on
Au)}
 \label{fig9}
\end{center}
\end{figure}

Integrating this spectrum over $Q^2$, we obtain (the upper
limit of this integration can be set to be equal to
infinity in a good approximation, due to the fast
convergence of the integral at $Q^2>1/R^2$):
\begin{equation}
\label{27}
dn_\gamma(W^2) = \frac{Z^2 \alpha }{ \pi} \, f\!\left(\frac{W^2R}{2
\omega}\right)\, \frac{dW^2}{W^2}\,.
\end{equation}
The function
\begin{equation}
\label{28}
f(x)=\int_{x^2}^{\infty}
\left(1-\frac{x^2}{y}\right)\,
F^2\!\left(\frac{y}{R^2}\right)\,
\frac{dy}{y}
\end{equation}
is presented in Fig.~\ref{fig8}. It is large for small
values of $x$,
\begin{equation}
\label{log1} f(x)= \ln\left(\frac{1}{ x^2}\right) - C_0
\qquad {\rm for} \qquad  x\ll 1 \,.
\end{equation}
(The value of constant $C_0$ depends slightly on the ratio
$a/R$: we obtain $C_0= 0.166$ for gold and $C_0= 0.163$ for
lead.) However, $f(x)$ drops very quickly for large $x$,
\begin{equation}
f(x) < \frac{1}{x^4} \qquad {\rm  for} \qquad x >  1 \,.
\label{30}
\end{equation}
Finally we obtain
\begin{equation}
\sigma^{\rm EPA}_{\rm Born}=\frac{Z^2 \alpha}{ \pi}
\int_{4m^2}^\infty\,\frac{dW^2 }{ W^2}\,
f\!\left(\frac{W^2R}{2\omega}\right)\,
\sigma_{\gamma \gamma} (W^2)\,.
\label{A6}
\end{equation}
A comparison of this cross section with the exact result is
shown in Fig.~\ref{fig9}. It is seen that the EPA gives an
accuracy better than 1~\% already at $\omega > 1.3$ GeV.
This needs to be explained.

%
%

Going from the exact expressions (\ref{39}), (\ref{sT}),
(\ref{sS}), (\ref{numbers}) to the approximate ones (\ref{sEPA}),
(\ref{gg}), (\ref{27}) we omit terms  of the relative order of
$Q^2/W^2$, which are dropped before the integration over $Q^2$ is
done. After the integration with the ``weight function''
$F^2(Q^2)/Q^2$ the
relative value of these corrections becomes of the order of
$1/(R^2W^2)$. In addition, the contribution of these correction
terms is suppressed by a logarithmic factor. Indeed, the main
contribution to the cross section in EPA is proportional to the
large Weizs\"acker-Williams logarithm
\begin{equation}
\label{LWW}
L_{\rm WW}= \int^{1/R^2}_{Q^2_{\min}}\frac{dQ^2}{Q^2}\approx
2\,\ln\left({\frac{\omega}{2m^2R}}\right)\,,
\end{equation}
while the omitted terms have no such a logarithm.
Therefore, the actual parameter describing the suppression
of the omitted terms to the differential cross section for
pair production is numerically small indeed,
\begin{equation}
\eta_{\rm EPA} \sim \frac{1}{R^2 \, W^2 \, L_{\rm WW}}\,.
\label{corEPA}
\end{equation}
%

%
%
\subsection{Simplified form factor: EPA}

The replacement of the realistic by the simplified form
factor means that we have to replace the function $f$ from
Eq.~(\ref{28}) by a function $\tilde f$ which is obtained
when we replace the form factor in the integrand in
Eq.~(\ref{28}) appropriately by the simplified nuclear form
factor. The SEPA (S here stands for the simplified form
factor) can thus be obtained using Eqs.~(\ref{sEPA}) and
(\ref{gg}) with the following expression for the number of
equivalent photons,
\begin{equation}
dn_\gamma(W^2) =
\frac{Z^2 \alpha }{ \pi} \,
\tilde{f}\!\left(\frac{W^2}{2 \omega\Lambda}\right)\,
\frac{dW^2}{W^2}\,.
\end{equation}
The function $\tilde{f}[W^2/(2 \omega\Lambda)]$ can be
obtained analytically,
\begin{equation}
\tilde{f}(\tilde{x})=(1 + 2\,\tilde{x}^2)\,
\ln{\left(\frac{1}{\tilde{x}^2} +1\right)}\, -\, 2 \,.
\label{fsimff}
\end{equation}
This is in contrast to $f(x)$, which would be the equivalent
of ${\tilde f}({\tilde x})$ for a realistic form factor
[see Eq.~(\ref{28})]. Now, ${\tilde f}({\tilde x})$ is
large for small values of $\tilde{x}$,
\begin{equation}
\label{log2}
\tilde{f}(\tilde{x}) \approx
\ln \left( \frac{1}{ \tilde{x}^2} \right)\, -\, 2
\qquad {\rm for} \qquad \tilde{x}\ll 1 \,,
\end{equation}
but drops very quickly for large $\tilde{x}$:
\begin{equation}
\tilde{f}(\tilde{x}) < \frac{1}{ 6\,\tilde{x}^4} \qquad
{\rm for}
\qquad \tilde{x} > 1 \,.
\end{equation}
Its behavior is presented by the dashed line in
Fig.~\ref{fig8}, where $x=R\Lambda\,\tilde{x}$.
In view of the same leading logarithmic asymptotics
for small argument [see Eqs.~(\ref{log1}) and~(\ref{log2})],
the functions $f$ and $\tilde{f}$ almost coincide for small
values of $x$.

Finally, we obtain for the simplified equivalent photon
approximation (SEPA),
\begin{align}
\sigma^{\rm SEPA}_{\rm Born} =& \;
\frac{Z^2 \alpha}{ \pi}
\int_{4m^2}^\infty\,\frac{dW^2 }{ W^2}\,
\tilde{f}\left(\frac{W^2}{ 2\omega \Lambda}\right)\,\sigma_{\gamma
\gamma} (W^2)
\nonumber\\[2ex]
=& \; \sigma_0\,J(\omega\Lambda/m^2)\,.
\label{A6s}
\end{align}
For large photon energies, the function $J(\omega\Lambda/m^2)$
behaves as
\begin{equation}
J(z)= \frac{28}{9}\,\left[ \ln\left(z \right) -
\frac{57}{14}\right]\qquad \mbox{for} \qquad z \gg 1\,.
\label{asymSEPA}
\end{equation}
A comparison of SEPA cross section with the exact result
from Eq.~(\ref{39}) is shown by the dashed and dot-dashed
lines in Fig.~\ref{fig9}. It is seen that the EPA gives a
considerable better accuracy than the SEPA, once again
confirming that the use of a realistic nuclear form factor
is essential.

%
%
\subsection{Realistic form factor: asymptotics for
the Born cross section}

\begin{figure}[htb]
\begin{center}
\includegraphics[width=1.0\linewidth]{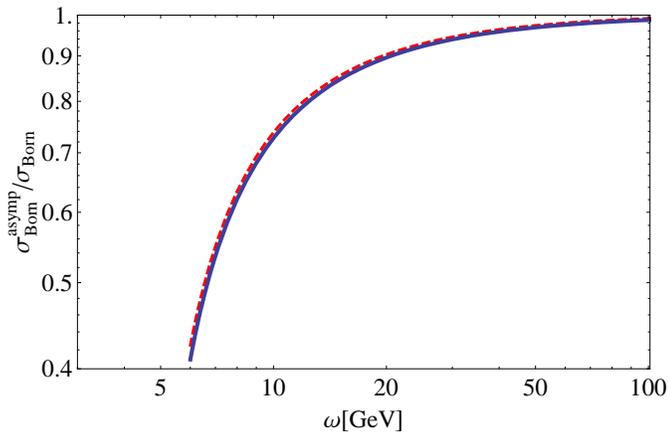}
\caption{Ratio $\sigma^{\rm asymp}_{\rm Born}/\sigma_{\rm
Born}$  for the realistic form factor (solid line)  and
ratio $\sigma^{\rm IM}_{\rm Born}/\sigma_{\rm Born}$
(dashed line) (photo-production on Au)}
 \label{fig10}
\end{center}
\end{figure}

The high-energy asymptotic behavior of the Born cross
section (for a realistic form factor) at large $\omega \gg
2m$ can easily be obtained using the EPA formulas
(\ref{sEPA}), (\ref{gg}), (\ref{27}) with the asymptotic
form of the function $f(x)$ given in Eq. (\ref{log1}). The
final result is
\begin{equation}
\sigma^{\rm asymp}_{\rm Born}= \frac{28}{9}\,
\sigma_0\,\left[
\ln\left(\frac{2\omega}{Rm^2}\right) - \frac{43}{ 14}-\frac 12 \, C_0
\right]\,.
\label{astrue}
\end{equation}
It should be noted that the cross section (\ref{astrue})
provides a reasonable approximation only for large enough
values of the photon energy $\omega$. Indeed, this cross
section is positive only at
\begin{equation}
\omega >\omega_{\rm crit} = \frac{1}{2}\,Rm^2\,
\exp{\left(\frac{43}{14} + \frac 12 C_0\right)} =
4.4\;\mbox{GeV}\,. \label{crit}
\end{equation}
A comparison of the asymptotics with the exact Born cross
section is given in Fig.~\ref{fig10}. It is seen that the
accuracy of a simple expression (\ref{astrue}) is better
than 10~\% only at very large $\omega > 20$~GeV, showing that the
realm of applicability of the high-energy asymptotics is
limited.

%
%

%
%
\subsection{Simplified form factor: result of Ivanov and Melnikov for asymptotics}

The cross section $\sigma_{\gamma Z}$ in the high-energy
limit was calculated by Ivanov and Melnikov in
Ref.~\cite{IM98} using the same expression (\ref{26}) for
the form factor of the nucleus and assuming
$\Lambda^2/(2m)^2 \ll 1$. The corresponding analytical
formula including the first correction $\sim
\Lambda^2/(2m)^2$ reads
\begin{align}
\sigma^{\rm IM}_{\gamma Z} =& \;
\sigma^{\rm IM}_{\rm Born}+\sigma^{\rm IM}_{\rm Coul},
\\[2ex]
\sigma^{\rm IM}_{\rm Born} =& \; \frac{28}{9}\,
\sigma_0\,\left[ \ln\left( \frac{2\omega \Lambda}{m^2} \right) -
\frac{57}{14} - C_1\right] \,,\;
 \label{A2B}
\\[2ex]
\sigma^{\rm IM}_{\rm Coul} =& \; -\frac{28}{9}\, \sigma_0\, C_2,
\label{A2}
\end{align}
where
\begin{equation}
C_1= \frac{12}{ 35} \left(\frac{\Lambda}{2m}\right)^2\,,
\quad C_2= 0.928\, (Z\alpha)^2\, C_1\,.
\label{A3}
\end{equation}
We note that the parameter $\Lambda^2/(2m)^2 =0.14$ is
small for muon pairs. A comparison of $\sigma^{\rm IM}_{\rm
Born}$ with the exact Born cross section (\ref{39}) is
shown by dashed line in Fig.~\ref{fig10}.

Two final remarks: {\it (i)} the SEPA asymptotics
(\ref{asymSEPA}) is in accordance with the result of Ivanov
and Melnikov~(\ref{A2B}), as has already been noted
in~\cite{IM98}. {\it (ii)} The difference between the
high-energy asymptotics $\sigma^{\rm asymp}_{\rm Born}$ for
the realistic form factor (\ref{astrue}) as opposed to the
high-energy asymptotics $\sigma^{\rm IM}_{\rm Born}$ for a
simplified form factor is very small:
\begin{equation}
\sigma^{\rm IM}_{\rm Born}-\sigma^{\rm asymp}_{\rm Born}=
0.012\; \frac{28}{9}\, \sigma_0\,.
\end{equation}
This is not surprising because the asymptotics are
determined
by a region with small values of $x=W^2R/(2\omega)$,
in which the spectra of the equivalent photons for the
realistic and simplified form factors coincide (see
Fig.~\ref{fig8}).

%
%
\section{Coulomb correction to the photo--production of pairs}
\label{section4}

Having discussed the role of the nuclear form factor in the
determination of the lepton pair production amplitude in
the Born approximation, we now turn our attention to the
role of Coulomb corrections. This is done according to our
``master equation''~(\ref{11master}). The Coulomb
correction is the leading correction beyond the Born
amplitude, provided the latter is being evaluated with
exact form factors.

\begin{figure}[htb]
\begin{center}
\includegraphics[width=1.0\linewidth]{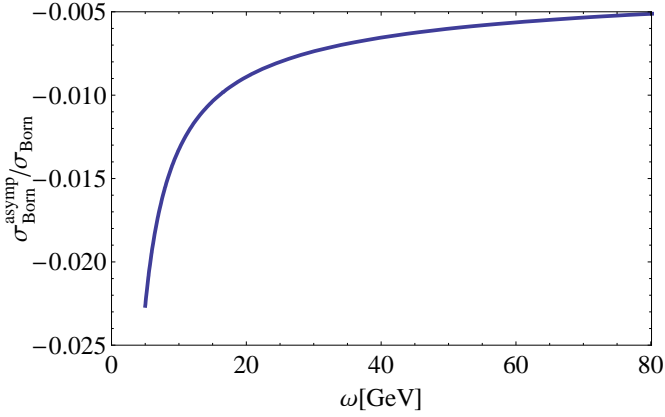}
\caption{Relative magnitude of the Coulomb correction
(photo-production on Au)}
 \label{fig11}
\end{center}
\end{figure}

The Coulomb correction corresponds to Feynman diagram of
Fig.~\ref{fig6}. The calculation of the Coulomb correction for
high photon energies ($\omega \gg 2m$) can be performed
approximately using the result of Ivanov and Melnikov given in
Eq.~(\ref{A2}). The ratio $\sigma^{\rm IM}_{\rm
Coul}/\sigma_{\rm Born}$ as presented at Fig~\ref{fig11} is
small. It is seen that the relative magnitude of the
Coulomb correction is less than 1~\% at $\omega > 20$ GeV.
This is in accordance with the following
estimate~\cite{IM98,HKS-2007}.

%
%

Due to the restriction of the transverse momenta of
additionally exchanged photons to the range below
$\Lambda\sim 1/R$, the effective parameter of the
perturbation series is not $(Z\alpha)^2$ but
\begin{equation}
(Z\alpha)^2\,\frac{\Lambda^2}{W^2}  \,.
\end{equation}
where $W$ is the invariant mass of the muon pair. Besides,
there is an additional logarithmic suppression because the
Coulomb corrections lack the large Weizs\"{a}cker--Williams
logarithm. Therefore, the actual parameter describing the
relative value of the Coulomb correction is
\begin{equation}
\eta_{\rm Coul}= (Z\alpha)^2\,\frac{\Lambda^2}{W^2\,
L_{\rm WW}}
 \label{corCoul}
\end{equation}
which corresponds to Coulomb corrections of less then
$1\,$\% for $\omega > 20$~GeV [we recall that $L_{\rm WW}$
is defined in Eq.~(\ref{LWW})].
It is reassuring that the
result of Ivanov and Melnikov confirms this estimate.
Indeed the relative order of the Coulomb correction
according to Eqs.~(\ref{A2B})---(\ref{A3}) is
\begin{align}
\frac{\sigma^{\rm IM}_{\rm Coul}}%
{\sigma^{\rm IM}_{\rm Born}} =& \;
0.318 \, (Z\alpha)^2\, \left(
\frac{\Lambda}{2m} \right)^2 \,
\nonumber\\[2ex]
\times & \; \left[  \ln\left(
\frac{2\omega \Lambda}{m^2} \right) - \frac{57}{14}-C_1
\right]^{-1}
\sim \eta_{\rm Coul}\,.
\end{align}

%
%
\section{Predictions for the RHIC and LHC colliders}
\label{section5}

We now turn our attention to the muon pair production in
collisions of heavy nuclei. Let us therefore consider the
process
\begin{equation}
Z(P_1)+Z(P_2) \to \mu^+(p_+)+\mu^-(p_-)+ Z(P_1')+Z(P_2')\,.
\label{procZZ}
\end{equation}
Its cross section can be calculated with a high accuracy by
means of the EPA using the result ~(\ref{39}) for the exact
cross section of the process,
\begin{equation}
\gamma(k)+ Z(P_2) \to \mu^+(p_+)+\mu^-(p_-)+ Z(P_2')\,.
\label{progZ}
\end{equation}
For the RHIC collider, we use the parameters $Z=79$ and
$\gamma=100$, the latter in order to be accordance with
the value used in
Ref.~\cite{Baltz-2009}. In the Born (B) approximation and
with a realistic (Fermi, F) form factor, we have:
\begin{equation}
\label{45}
\sigma_{\rm BF}^{ZZ}=
\frac{Z^2\alpha}{\pi}\int_{2m}^{\infty}\frac{d\omega}{\omega}
f\left(\frac{\omega R}{\gamma_L}\right) \, \sigma_{\rm
Born}^{\gamma Z}(\omega)= 0.193\;\mbox{barn} \,.
\end{equation}
In Eq.~(\ref{45}),
$\gamma_L=2\gamma^2$ is the Lorentz-factor of the first
nucleus in the rest frame of the second nucleus; $f(x)$ and
$\sigma_{\rm Born}^{\gamma Z}(\omega)$ can be found in
Eqs.~(\ref{28}) and (\ref{39}), respectively. There is a
$9.8$\% difference to the corresponding result for the
simplified (S) form factor, still in the first Born
approximation,
\begin{equation}
\sigma_{\rm BS}^{ZZ}=0.212\;\mbox{barn}\,.
\end{equation}
This is in full agreement with the recent result $0.211$
barn of Ref.~\cite{Baltz-2009}. The consistent use of
$\Lambda = 80$ is crucial in order to obtain this
agreement. We note in passing that $\Lambda =90$~MeV
results in a 22~\% difference.

A calculation for the Coulomb correction in LLA can be done
using the result of Ivanov-Melnikov and taking into account
Coulomb corrections to both nuclear lines (factor~2),
\begin{equation}
\sigma_{\rm Coul}^{ZZ}=
\frac{Z^2\alpha}{\pi}\int_{2m}^{\infty} \frac{d\omega}{\omega}
f\left(\frac{\omega R}{\gamma_L}\right) \,2\sigma_{\rm
Coul}^{\rm IM}= -0.0072 \;\mbox{barn}\,.
\end{equation}
It means that the relative value of the Coulomb correction
is $-3.7$~\% in full contrast to the recent result $-22$~\%
of Ref.~\cite{Baltz-2009}, but in agreement with our
parametric estimates.

For the LHC collider, we use $Z=82, \gamma=2760$, again in
order to be in accordance with Ref.~\cite{Baltz-2009}. We
have for a realistic form factor,
\begin{equation}
\label{46}
\sigma_{\rm BF}^{ZZ}= 2.36\;\mbox{barn} \,,
\end{equation}
and for a simplified form factor,
\begin{equation}
\sigma_{\rm BS}^{ZZ}=2.45\;\mbox{barn}\,.
\end{equation}
This is in good agreement with the recent result $2.42$
barn of Baltz~\cite{Baltz-2009}. Again, an estimate for the
Coulomb correction can be obtained on the basis on an
integration over the result of Ivanov and Melnikov,
\begin{equation}
\sigma_{\rm Coul}^{ZZ}= -0.03 \;\mbox{barn}\,.
\end{equation}
It means that the relative value of the Coulomb correction
is $-1.3$~\% in full contrast to the recent result $-14$~\%
of Baltz~\cite{Baltz-2009}.

For completeness, we recall that in Table~\ref{t1},
slightly different values were used for the relativistic
Lorentz factors at the modern colliders, namely,
$\gamma=108$ (RHIC) and $\gamma = 3000$ (LHC)
instead of
$\gamma=100$ (RHIC) and $\gamma = 2760$ (LHC).
In both cases, with the alternative values for $\gamma$
and for a realistic nuclear form factor,
we obtain results for $\sigma_{\rm BF}^{ZZ}$
which are slightly larger than those in Eqs.~(\ref{45})
and~(\ref{46}), namely $0.209\;\mbox{barn}$ for RHIC
and $2.46\;\mbox{barn}$ for the LHC (see Table~\ref{t1}).

%
%
\section{Conclusions}
\label{section6}

We have analyzed in detail the role of the nuclear form
factor in the calculation of muon pair production cross
sections in photon-nucleus and nucleus-nucleus collisions.
At RHIC, the realistic (Fermi) nuclear charge distribution
leads to predictions that deviate by $10\div20$\% from the
corresponding values for simplified nuclear form factors.
We also show quantitatively that  the EPA is an
excellent approximation to the muon photo-production for
photon energies that exceed the rest mass of the produced
pair (region $\omega\gg 2m$) as well as for muon pair
production at RHIC and LHC.

We find that the Coulomb corrections for the muon
production are less pronounced than for the $e^+e^-$ pair
production. Our calculation in LLA leads to a decrease by
about $1.3\div 3.7$~\% due to higher-order Coulomb effects
at the LHC and RHIC colliders.

Let us issue a few remarks regarding the obvious
discrepancy of our results about the Coulomb corrections to
those of the recent, interesting paper~\cite{Baltz-2009}.
It is not obvious from the condensed presentation given in
Ref.~\cite{Baltz-2009} whether or not the nuclear form
factors have been taken into account to all orders in
$Z\alpha$. Therefore, the approach may need to be
reexamined. Moreover, our parametric quantitative estimates
given by the numerically small expansion parameter
(\ref{corCoul}) indicate that the Coulomb corrections for
muon pair production should be smaller than those for
$e^+e^-$ production. Coulomb corrections
for the total production cross section of
heavier lepton pairs would be even smaller,
and in addition, we note that the Coulomb
corrections also decrease with
higher invariant mass $W^2$. Under typical conditions,
muons from the discussed
process can be detected at the RHIC and LHC colliders
with large values of $W$; this  means that the Born
approximation can be safely used in numerical simulations
of this process. A correction on the order of 22~\% for
muons at RHIC, as obtained in Ref.~\cite{Baltz-2009}, is
larger than that for the $e^+e^-$ production and seems
unrealistically large even if we allow for a large
numerical prefactor multiplying the parameter
(\ref{corCoul}). We also note that
the calculation of the Coulomb
corrections in Ref.~\cite{Baltz-2009} proceeds in the
impact parameter representation. The numerical evaluation
of integrals of this type is known to be notoriously
problematic because of large numerical
cancellations due to oscillations. In
any case, a calculation of the Coulomb corrections beyond
the leading logarithmic approximation is desirable.

Finally, it should be mentioned than unitarity corrections
to the muon production have been discussed in
Ref.~\cite{HKS-2007,JHS-2008} with the following result:
Unitarity corrections for the exclusive production of
exactly one muon pair are large. However, the experimental
study of the exclusive muon pair production seems to be a
very difficult task, because it requires that the muon pair
should be registered without any electron--positron pair
production, including $e^\pm$ emitted at very small angles.
The corresponding inclusive cross section is not affected
by the unitarity correction and, indeed,
close to the Born cross section.

\section*{Acknowledgments}

We are grateful to D.I.~Ivanov, R.N.~Lee, A.I.~Milstein and
V.N.~Pozdnyakov for useful discussions. U.D.J. acknowledges
support from the National Science Foundation (PHY-8555454)
and from the Missouri Research Board. V.G.S.~is supported
by the Russian Foundation for Basic Research via grant
09-02-00263 and acknowledges support from GSI Darmstadt
(project HD--JENT).


\end{document}